\documentclass{b98proc}

\def\Journal#1#2#3#4{{#1} {\bf #2}, #3 (#4)}

\def\NPA{{\em Nucl. Phys.} A}

\def\PLB{{\em Phys. Lett.} B}

\def\PRL{\em Phys. Rev. Lett.}
\def\PREV{\em Phys. Rev.}

\def\PRD{{\em Phys. Rev.} D}
\def\PRC{{\em Phys. Rev.} C}

\def\beq{\begin{equation}}
\def\eeq{\end{equation}}
\def\beqn{\begin{eqnarray}}
\def\eeqn{\end{eqnarray}}

\begin{document}

\title{
\vspace*{-0.5cm}
\hfill{\rm MKPH-T-98-20}\\
\vspace*{0.5cm}
FORM FACTORS AND EXCLUSIVE PROCESSES -- INTRODUCTION AND OVERVIEW
}

\author{D. DRECHSEL} \address{J. Gutenberg-Universit\"at, Institut f\"ur
  Kernphysik, 55099 Mainz, Germany}


\maketitle

\abstracts{ Form factors, polarizabilities and excitation spectra are
  different aspects of many-body structures. These aspects are
  constrained by sum rules based on general principles like
  relativity, gauge invariance, causality and unitarity. Several such
  structure effects of baryons and mesons will be reviewed, with
  particular emphasis on the neutron charge form factor, the
  strangeness content of the nucleon, polarizabilities, and the
  Gerasimov-Drell-Hearn and related sum rules.  }

\section{Introduction}
Hadrons are complex systems with many internal degrees
of freedom. In principle, their constituents are current quarks and
gluons interacting by the laws of QCD. Due to these laws the colour
interaction increases with decreasing energy and momentum involved,
and as a consequence hadrons exist as highly correlated many-body
systems. As in any such system, it becomes important to find the
proper ``effective'' degrees of freedom, which in this case are the
low-energy realizations of quarks and gluons, the colourless mesons
and baryons.

Internal degrees of freedom have two immediate consequences, a finite
size of the object and an excitation spectrum. In the case of the
nucleon the finite size effect may be described by the electric (E)
and magnetic (M) Sachs form factors,

\beqn G_E^N(Q^2)& = & e_N-\frac{Q^2}{6}<r^2>_E^N+[Q^4] \ ,\nonumber \\ 
G_M^N(Q^2)& = &(e_N+\kappa_N) \left
  (1-\frac{Q^2}{6}<r^2>_M^N+[Q^4]\right ) \ ,
\label{DDeq1.1}
\eeqn

\noindent
with normalizations $e_N=1$ or 0 for $N$=proton or neutron
respectively, and $\kappa_N$ the anomalous magnetic moment of the
nucleon. The form factors are functions of 4-momentum transfer,
$q^2=(k_f-k_i)^2=-Q^2$, where $k_f$ and $k_i$ are the 4-momenta of the
leptons in the final and initial states, respectively.  The quantity
$Q^2$ has been defined to be positive for lepton scattering, which
leads to the exchange of a ``space-like'' virtual photon $(q^2<0)$.
Pair production, on the other side, requires the exchange of a
``time-like'' virtual photon and becomes physically possible for
$q^2\geq 4m_N^2$, whenever the mass $2m_N$ of the nucleon-antinucleon
pair can be produced. The mathematically most interesting part of the
form factors lies in the unphysical region $0<q^2<4m_N^2$, where the
function is strongly increasing and rapidly fluctuating due to the
vicinity of the vector meson poles in the complex plane.

A particle with an anomalous magnetic moment can only exist if it has
a finite size at the same time, elsewise one would encounter serious
divergencies in the high-energy limit. Similarly the existence of a
finite size has the consequence that the particle can be polarized by
a (quasi)static electromagnetic field. A physical process to determine
the polarizabilities of the nucleon is Compton scattering of photons
with sufficiently low energy $\omega$.  The leading term of the
Compton amplitude is the Thomson term, which depends only on
``global'' properties of the target.  The subleading term of order
$\omega^2$ is proportional to the electric $(\alpha)$ and magnetic
$(\beta)$ polarizabilities describing the deformation under the
influence of an external electromagnetic field. For the neutron the
expansion starts with the second term. The corresponding cross section
is given by the square of the scattering amplitude, i.e. it scales
with $\omega^4$. This is the case of Rayleigh scattering, also
observed in the scattering of light off neutral atoms leading to the
blue sky.

We can visualize the relation between the polarizability and Compton
scattering $h(\gamma, \gamma')h'$ as follows: The incoming photon
$\gamma$ polarizes the hadron $h$ whose charges arrange in an
energetically more favourable distribution. The new equilibrium is
described by a coherent superposition of ground and excited state wave
functions, which is subsequently analyzed by studying the angular
distribution of the outgoing photon.

Recently it has also become possible to observe so-called generalized
polarizabilities in the reaction $p(e,e'\gamma )p'$ with virtual
incident photons. These generalized polarizabilities depend on the
4-momentum transfer, i.e. $\alpha=\alpha (Q^2)$, and their Fourier
transforms describe the spatial distribution of the polarization
densities.

The reaction $\vec{e}+\vec{N}\rightarrow$ anything is parametrized by
4 response functions $\sigma_T$, $\sigma_L$, $\sigma_{TT}'$ and
$\sigma_{LT}'$, which are functions of energy and momentum transfer.
The indices $T$ and $L$ refer to the transverse and longitudinal
polarization of the virtual photon, and the latter two structure
functions can only be studied by double-polarization experiments.
Several energy-weighted integrals over these structure functions are
related to static and quasistatic properties of the hadronic system.
These ``sum rules'' are based on Lorentz and gauge invariance,
analyticity and unitarity of the scattering amplitudes. They can
be derived by combining the results of low-energy theorems with
dispersion relations and the optical theorem. Typical examples are the
sum rules of Baldin, Gerasimov-Drell-Hearn (GDH), Burkhardt-Cottingham
(BC) and Bjorken.

\section{Form Factors}

The form factors of the nucleon, in particular the evasive charge form
factor of the neutron, are now being studied by double polarization
experiments, $\vec{e}+\vec{N} \rightarrow e'+N'\ \ \mbox{or}\ \ 
\vec{e}+N \rightarrow e'+\vec{N}'\ .$ The differential cross section
for these reactions takes the form~\cite{DDArn81}

\beqn
\label{DDeq2.2}
\frac{d\sigma}{d\Omega} = \underbrace{\cdots (G_E^{\,2}+\cdots G_M^{\,2})}_
                          {(d\sigma/d\Omega)_{unpol.}}+
                          \underbrace{\cdots P_eP_N^{\perp}G_EG_M}_
                          {A_{\perp}}+
                          \underbrace{\cdots P_eP_N^{\|}G_M^{\,2}}_
                          {A_{\|}}\ .
\eeqn

\noindent
The ellipses in this equation indicate known kinematical factors.
Since the Sachs form factors in the first term add in squares, it
is difficult if not impossible to determine the (small) charge form
factor of the neutron by an unpolarized experiment. The last two terms
can be measured with longitudinally polarized incident electrons
(polarization $P_e$) and polarized targets (or recoil polarization),
$P_N^{\perp}$ and $P_N^{\|}$ referring to polarizations perpendicular
and parallel to the momentum transfer.  Except for known kinematical
factors, the ratio of the asymmetries, $A_{\perp}/A_{\|}$, is
proportional to $G_E/G_M$, i.e. $G_E$ can be determined in a
model-independent way once $G_M$ is known.  The observable $G_E^n$ was
recently studied at MAMI by means of two different experiments,
scattering of polarized electrons off a polarized $^3He$ target, and
off an unpolarized $^2H$ target but analyzing the recoil polarization
of the ejected neutron. The results will be presented by H. Schmieden
and P. Grabmayr~\cite{DDSch98}. There are indications for two
surprises:

\begin{itemize}
\item The reaction was previously assumed to be quite independent of
  final-state interactions, at least in quasifree kinematics and for
  perpendicular polarization. It has now been shown that in the case
  of $^2H$ the corrections increase dramatically for smaller momentum
  transfer. Therefore, the observed discrepancy between the
  $^3He$ and $^2H$ results could well be due to an uncomplete
  treatment of final-state interactions and two-body currents in the
  case of $^3He$.
\item All models (Nambu-Jona-Lasinio, solitons, constituent quarks,
  chiral and cloudy bags, etc.) predict a core with positive charge
  inside and a negative cloud outside.  However, the predictions
  differ substantially in absolute value and zero-crossing of the
  charge distribution. The recent deuteron data indicate a
  zero-crossing at a small radius of about 0.7 fm, close to the axial
  radius, which is determined by the quark bag radius in chiral bag
  models. The previously preferred ``Platchkov fit'' had the
  zero-crossing at about 0.9 fm, close to the proton radius. At the
  same time the values for $G_E^n$ and $|\rho_E^n|$ increase by as much as
  a factor of~2.
\end{itemize}

\noindent
R. Baldini~\cite{DDBal98} will report on the FENICE experiment
$e^+e^-\rightarrow n\bar{n}$, which probes the neutron form factors in
the time-like region. The main results are the following:

\begin{itemize}
\item The time-like form factors are at about twice the level of the
  space-like ones at the same values of $|q^2|$.  This is a clear sign
  that asymptotia will only be reached at much higher momentum
  transfer.
\item The neutron form factors at $q^2>0$ are about twice the value of
  the proton form factors, i.e. there is considerable absorptive
  strength of isovector-vector mesons in and above the threshold
  region.
\item There are strong indications of a resonance structure near
  threshold, and some indications that both $G_E^n$ and $G_M^n$ could
  become small in this region (note: $G_E(4m^2)=G_M(4m^2)$ by
  definition).
\end{itemize}

\noindent
The finite size of the nucleon is also addressed in the following
contributions:

\begin{itemize}
\item The surprising results of the FENICE experiment are described by
  S. Dubnicka, A.-Z. Dubnickov\'{a} et al.~\cite{DDDub98} by a new
  dispersion fit to the data.  In particular this fit needs a
  $\rho^{IV}(2.5$ GeV) with large width.
\item R. Bijker and A. Leviatan have calculated the form factors in
  the framework of the algebraic model developed in collaboration with
  Iachello, and S. Boffi will present the results of the chiral
  constituent quark model~\cite{DDBij98}. In both cases the data
  require an intrinsic quark form factor corresponding to a quark
  radius of about 0.6 fm, which is close to the radius of the wave
  function. This finite-size constituent quark can be visualized as a
  point-like (current) quark surrounded by a pion cloud. However,
  there are two immediate consequences: The virtual pion surrounding
  the quark can become real whenever sufficient energy is supplied,
  i.e. the quark size will also influence the excitation spectrum, and
  the crowded environment of large constituent quarks in a small bag
  will create strong exchange currents, leading to sizeable
  corrections for all observables as shown by G.~Wagner and
  A.~Buchmann~\cite{DDWag98}.

\item H.~Haberzettl et al.~\cite{DDHab98} have shown the need to
  include finite size effects for a consistent description of meson
  photoproduction.
\item I. Eschrich~\cite{DDEsc98} will present the latest results of
  the SELEX collaboration at Fermilab, a pion radius in good agreement
  with previous experiments, and for the $\Sigma^-$ a value of
  $<r^2>_{\Sigma^-}=(0.60\pm 0.08\pm 0.08)$\ fm$^2$, markedly smaller
  than for the proton.
\end{itemize}

\section{Strangeness}
A sizeable strange sea in the nucleon is predicted from both the
$\sigma$ term (extracted from pion-nucleon scattering) and deep
inelastic lepton scattering (DIS). New and independent information can
be obtained from parity-violating electron scattering,
$N(\vec{e},e')N'$, by studying the interference term of virtual photon
$(\gamma^{\ast})$ and $Z^{\circ}$ exchange. Since the $Z^{\circ}$ can
couple to both vector and axial vector currents, part of this
interference term changes sign under an helicity flip of the incident
electron. The resulting asymmetry is of order ${\cal A}\approx
10^{-4}Q^2/$GeV$^2$ and provides information on the electric
$(\tilde{G}_E)$, magnetic $(\tilde{G}_M)$ and axial $(\tilde{G}_A)$
form factors as seen by the $Z^{\circ}$.

Within the standard model and assuming isospin symmetry, these
additional form factors determine the contributions of u, d and s
quarks,

\begin{eqnarray}
\gamma^{\ast}p \rightarrow G^p
     & = &\frac{2}{3}G_u-\frac{1}{3}G_d-\frac{1}{3}G_s \nonumber\\
\gamma^{\ast}n \rightarrow G^n
     & = &\frac{2}{3}G_d-\frac{1}{3}G_u-\frac{1}{3}G_s\nonumber \\
Z^{\circ}p \rightarrow \tilde{G}^p
     & = & \left (\frac{1}{4}-\frac{2}{3}\sin^2\theta_W\right )G_u+
           \left (-\frac{1}{4}+\frac{1}{3}\sin^2\theta_W\right )
           (G_d+G_s)\ ,
\label{DDeq3.1}          
\end{eqnarray}

\noindent
with $\theta_W$ the Weinberg angle and $\sin^2\theta_W\approx 0.2325$.

The pioneering experiment was performed in 1997 at
MIT/Bates~\cite{DDMue97}. Much to the surprise of most theorists, the
experiment yielded a positive value of the strange magnetic
moment, however with large error bars.  In April and May this year,
also the HAPPEX collaboration at Jefferson Lab obtained a first result
for the asymmetry.  Another experiment is being prepared at MAMI by
the A4 collaboration.

The status of the field will be reviewed by M. Pitt, and R.~Michaels
will report on HAPPEX~\cite{DDPit98}. The latter experiment was
performed at $Q^2=0.48$ GeV$^2$ and yielded an asymmetry ${\cal
  A}_{\rm raw}=-5.64\pm 0.75$ in ppm.  However, most of this asymmetry
is due to u and d quarks, and the result for the strange
sea is still compatible with zero, $G_E^{\,s}+0.39 G_M^{\,s} =
0.019\pm 0.035\pm 0.023\pm 0.026,$ with errors due to the statistics,
systematical uncertainties, and the presently bad knowledge of the
neutron form factor.

In the framework of chiral perturbation theory, $G_M^s(0)=\mu_p^s$
cannot be predicted, because it involves an unknown low-energy
constant. However, T.  Hemmert~\cite{DDHem98} will report on a recent
calculation in HBChPT that determines the slope of $G_M^s$ at $Q^2=0$.
For the central value of the SAMPLE experiment, the extrapolation
leads to $0.03\leq \mu_p^s\leq 0.18$.

\section{Polarizabilities}
The electric $(\alpha)$ and magnetic $(\beta)$ polarizabilities of a
macroscopic system describe its response to external quasistatic
electric $(\vec{E})$ and magnetic $(\vec{H})$ fields. The system
rearranges its charge and magnetization distributions, which results
in a lowering of the total energy,

\beq
\Delta E = -\frac{1}{2} \alpha \vec{E}^2-\frac{1}{2}\beta\vec{H}^2\ .
\label{DDeq4.1}
\eeq

\noindent
For a metal sphere the electric polarizability $\alpha$ is essentially
given by the volume $V$, because the charges can easily move to the
surface. For a dielectric sphere an additional factor ($\epsilon -1$)
appears, and since the dielectric constant $\epsilon$ is close to $1$,
$\alpha$ is substantially reduced.  In a similar way, the nucleon is a
very rigid object. It is difficult to deform its charge distribution,
which leads to a small ratio $\alpha/V \approx 2 \cdot 10^{-4}$.

Due to its spin, the nucleon has 4 additional vector (or spin)
polarizabilities labeled $\gamma_1$ to $\gamma_4$, in addition to the
2 scalar polarizabilities $\alpha$ and $\beta$. These 6 observables
can be measured in Compton scattering by determining the low energy
expansion of 6 independent Compton amplitudes. In particular the
forward Compton amplitude takes the form

\beq
T (\omega, \theta) = \hat{\epsilon}' \cdot \hat{\epsilon}
f (\omega) + i \vec{\sigma} \cdot (\hat{\epsilon}' \times \hat{\epsilon})
g (\omega),
\label{DDeq5.1}
\eeq

\noindent
where $\vec{\sigma}$ is the spin of the nucleon, and $\hat{\epsilon}$
and $\hat{\epsilon}'$ are the polarizations of the photons in the
initial and final states, respectively. Since the amplitude has to
obey the crossing symmetry, $f(\omega)$ is an even function of
$\omega$ and $g (\omega)$ is odd.

As has been stated before, the leading term in $\omega$ is given by
static properties of the target, and gauge invariance determines the
subleading term in a model-independent way. The dynamical properties
of the system appear at relative order $\omega^2$, and are
parametrized by the polarizabilities,

\begin{eqnarray}
\label{DDeq4.3}
f(\omega) & = &-\frac{e^2}{m}+4\pi (\alpha +\beta)\ \omega^2+[\omega^4]\ , \\
g(\omega) & = & -\frac{e^2\kappa^2\omega}{2m^2}+4\pi \nonumber
                       (\gamma_1-\gamma_2-2\gamma_4)\ \omega^3+\cdots\ .
\end{eqnarray}

\noindent
In order to determine all 6 polarizabilities independently, it will be
necessary to perform double polarization experiments such as
$\vec{\gamma} + \vec{p} \rightarrow \gamma' + p'$. Of particular
current interest are the forward and backward spin polarizabilities,
given by $\gamma_0 = \gamma_1 - \gamma_2 - 2 \gamma_4$ and
$\gamma_{\pi} = \gamma_1 + \gamma_2 + 2 \gamma_4$, respectively. It
has been shown that $\gamma_0$ is determined by the difference of
s-wave pion production (multipole $E_{0^+}$) and $\Delta$(1332)
excitation (multipole $M_{1+}$), $\gamma_0 \approx 2.5$ (s-wave) $-3.0
(\Delta) - 0.1 \approx - 0.6$ (here and in the following in units of
$10^{-4}$fm$^4$)~\cite{DDDre98a}.  While it is difficult to measure
$\gamma_0$ by Compton scattering, it can be determined from the
exclusive structure function $\sigma_{TT}'$ by a sum rule (see chapter
5). From a combined analysis of (unpolarized) Compton scattering and
pion photoproduction, Tonnison et al. have found a value of
$\gamma_{\pi} = - 27.1 \pm 3.4$~\cite{DDTon98}.  The theoretical
predictions, however, are in the range of $-40 < \gamma_{\pi} < - 34$,
the bulk contribution of about $-44$ due to the close-by pion pole
($\pi^0 \rightarrow 2 \gamma$, Wess-Zumino-Witten term or triangle
anomaly). Within the framework of backward dispersion relations, A.
L'vov and A. Nathan have included the $\pi N$ and $\pi \pi N$
contributions in the s channel, and $\pi^0, \eta, \eta'$ in the $t$
channel, and predicted $\gamma_{\pi} = -39.5$~\cite{DDLvo98}. The
puzzle should be solved by polarized Compton scattering. The asymmetry
defined by parallel vs. antiparallel spin projections of photon and
proton is very large and varies between values of $50\%$ and $80\%$,
depending on the size of $\gamma_{\pi}$~\cite{DDPas98}.

Virtual Compton scattering (VCS), $\gamma^{\ast}+p \rightarrow \gamma
+p'$, is realized by the reaction $e + p \rightarrow e' + p' +
\gamma$, with radiation from both the nucleon and the electron.  The
latter is the Bethe-Heither process determined by QED.  Due to the
small mass of the electron it is the dominant process, particularly in
the directions of the incident and scattered electrons. Assuming that
the information can be read off the interference term between the
Bethe-Heitler process and VCS, P.  Guichon~\cite{DDGui95} has
expressed the reaction in terms of generalized polarizabilities (GPs),
$\alpha(Q^2)$ etc. These can be obtained by coupling the transverse
electric, transverse magnetic and longitudinal multipoles of the
incident virtual photon with the transverse electric and magnetic
multipoles of the emitted real photon. There are 3 possibilities to
couple these multipoles to zero and 7 possibilities to obtain one,
resulting in 3 scalar and 7 vector GPs. Since there exist 4
model-independent relations between these GPs, only 6 GPs are
independent~\cite{DDDre97}. However, the unpolarized experiment yields
only 3 independent observables, and it will take polarization degrees
of freedom to determine all 6 GPs, e.g. the reaction
$\vec{e}+p\rightarrow e'+\vec{p}\ '+\gamma$ ~\cite{DDVan97}.

First results of a pilot experiment at MAMI will be shown by N.
d'Hose, and J. Friedrich will report on the experimental
details~\cite{DDdHo98}.  The experiment has been performed at fixed
momentum of the virtual photon and for real photons with momenta
$33.6$ MeV $<q'<111.5$ MeV. As should be expected, the cross sections
at low values of $q'$ are essentially determined by QED, but with
increasing $q'$ the GPs become visible. Since these effects
are of the order of $10^{-38}cm^2/MeV sr^2$ in the differential cross
section, the experiment requires a very careful analysis of higher
order QED corrections and systematical errors.

Finally, there are two contributions to the Conference regarding the
structure of the pion. L.V. Fil'kov and V.L. Kashevarov~\cite{DDFil98}
predict new values for the pion polarizabilities from dispersion
relations and discuss the possibility to measure these quantities by
the reaction $\gamma + p \rightarrow \pi^+ + \gamma' + n$ at MAMI.
Such an experiment was performed at the Lebedev Institute in the 70's
with the result $\alpha - \beta = 40 \pm 24$, while ChPT predicts
$5.3$ (in units of $10^{-4}$ fm$^3$). Similar and other problems occur
in the analysis of two other reactions, radiative pion scattering off
a heavy nucleus and pion pair annihilation.

As will be reported by M. A. Moinester~\cite{DDMoi98}, the SELEX/E781
collaboration at Fermilab is investigating the reactions $e \pi
\rightarrow e' \pi' \gamma$ and $e \pi \rightarrow e' \pi' \pi^{0}$
for incident pions of 600 GeV. The experiment is expected to provide
new information on the excitation spectrum of the pion, in particular
on the GPs of the pion and the chiral anomaly $\gamma^{\ast}
\rightarrow 3 \pi$.

\section{Sum Rules} 

Assuming that $f(\omega)$ is an analytical function as required by
causality, one can derive a once-subtracted dispersion relation,

\begin{eqnarray}
\label{DDeq5.2}
Re\ f(\omega) & = & f(0)+\frac{2\omega^2}{\pi}\int
                    \frac{\mbox{Im}\ f(\omega')}{\omega'(\omega'^2-\omega^2)}
                     d\omega'\\ \nonumber
              & = & f(0)+\frac{2}{\pi}\int
                    \frac{\sigma_{\rm{T}}(\omega')}{\omega'^2}\ d\omega'
                    \cdot \omega^2+[\omega^4]\ ,
\end{eqnarray}

\noindent
involving integration from one-pion threshold, $\omega' =
\omega_{thr}$, to infinity. The second line of Eq.~(\ref{DDeq5.2})
follows from unitarity of the $S$ matrix and a Taylor expansion in
$\omega$. By comparing this result with the LET, Eq.~(\ref{DDeq4.3}),
we find

\beq
\alpha + \beta = \frac{1}{2 \pi^2} \int \frac{\sigma_T (\omega)}
{\omega'^2} d \omega,
\label{DDeq5.3}
\eeq

\noindent
which is Baldin's sum rule~\cite{DDBal60}. Of course, the power series
of Eq.~(\ref{DDeq5.2}) converges only for $\omega < \omega_{thr}$, for
larger values of $\omega$ the function turns complex and has an
infinite number of branch cuts due to multiple particle production.

The spin-flip amplitude $g$ can be studied by double polarization
experiments. The corresponding absorption cross section will be
denoted by $\sigma_{1/2}$ and $\sigma_{3/2}$ for antiparallel and
parallel spins of photon and target nucleon, respectively.  The
transverse responses are then given by $\sigma_T = (\sigma_{1/2} +
\sigma_{3/2})/2$ and $\sigma_{TT}' = (\sigma_{3/2} - \sigma_{1/2})/2$,
and the dispersion relation for $g (\omega)$ may be cast into the form

\begin{eqnarray}
\label{DDeq5.5}
Re\ g(\omega)&=& \frac{2\omega}{\pi}\int\frac{Im\ g(\omega')}
                    {\omega'^2-\omega^2}\ d\omega' \\
            & = & -\frac{2}{\pi}\int\frac{\sigma_{TT}'(\omega')}
                 {\omega'}\ d\omega'\cdot \omega - 
                 \frac{2}{\pi}\int\frac{\sigma_{TT}'(\omega')}
                 {\omega'^3}\ d\omega'\cdot \omega^3 + [\omega^5]\nonumber\ .
\end{eqnarray}

\noindent
Comparing again with Eq.~(\ref{DDeq4.3}), we obtain the relations

\beqn
I = \int \frac{\sigma_{1/2}(\omega)-
    \sigma_{3/2}(\omega)}{\omega} d\omega = -\frac{\pi e^2}{2m^2}\kappa^2
\label{DDeq5.6}
\eeqn

\noindent
and

\beqn
\gamma_0 =  \frac{1}{4\pi^2}\int\frac{\sigma_{1/2} (\omega) -\sigma_{3/2}
            (\omega)} {\omega^3}\ d\omega\ .
\label{DDeq5.7}
\eeqn

\noindent
The former relation is the famous GDH sum rule~\cite{DDGer65}, the
second one provides a recipe to calculate the forward spin
polarizability~\cite{DDGel54}.  While the pion photoproduction
multipoles add in squares in the case of the total absorption cross
section $\sigma_T (\omega)$, they carry alternating signs in the case
of the spin-flip amplitude,

\begin{eqnarray}
\label{DDeq5.8}
\sigma_{1/2}-\sigma_{3/2}& \sim & |E_{0+}|^2-|M_{1+}|^2+6E_{1+}^{\ast}M_{1+}+
                         3|E_{1+}|^2 \pm \ldots
\end{eqnarray}

\noindent
Therefore the sum rules of Eqs.~(\ref{DDeq5.6}) and (\ref{DDeq5.7})
are very sensitive to small changes in the individual multipoles. Due
to the weight factors $\omega^{-1}$ and $\omega^{-3}$, the $s$-wave
threshold amplitude is particularly enhanced.

In the past the GDH integral was evaluated by using the pion
photoproduction multipoles of Eq.~(\ref{DDeq5.8}) and some model
description for the more-pion channels. The first such calculation was
performed by Karliner~\cite{DDKar73}. In units of $\mu b$ she
predicted $I^p = - 261$ and $I^n = -183$, at variance with the GDH
result $I^p = -205$ and $I^n = -233$. The discrepancy is particularly
obvious for the isoscalar-isovector interference $I^p - I^n$, in which
case the prediction differs in sign and by a factor of 3. The
situation became even more puzzling when Sandorfi et
al.~\cite{DDSan94} evaluated the integral using the SAID multipoles.
The result was $I^p - I^n = -129$, to be compared with $+28$ according
to GDH. As was shown recently, however, a large fraction of the
difference is due to the fact that the $s$-wave multipoles used in the
calculation miss the threshold value of the Kroll-Ruderman theorem by
about 20\%~\cite{DDDre98b}.

The pioneering experiment on the helicity structure of photoabsorption
was recently performed at MAMI, and H.J. Arends~\cite{DDAre98} will
report on the preliminary results obtained by the GDH collaboration in
the energy range 200MeV$<$$\omega$$<$800 MeV. The data clearly
indicate an opposite sign of the $E_{0+}$ vs. $M_{1+}$ contributions
below the $\Delta$ resonance and an excess of two-pion production over
the one-pion prediction above the $\Delta$. The data collection will
be continued at ELSA in order to study whether and how the GDH
integral (weighting factor $\omega^{-1}$!) saturates at the higher
energies. However, already the present data should give a good value
for the forward pion polarizability $\gamma_0$ due to weighting factor
$\omega^{-3}$.

The sum rules and integrals for real photons can be generalized to the
case of virtual photons. The total virtual cross section $\sigma_{v}$
is given by a flux factor $\Gamma$ and the four response functions
mentioned above,

\begin{equation}
\label{DDeq5.9}
\sigma_v = \Gamma \left [\sigma_T+\varepsilon_L\sigma_L+
           P_eP_x\sqrt{2\varepsilon_L(1-\varepsilon)}\sigma'_{LT}+
           P_eP_z\sqrt{1-\varepsilon^2}\sigma'_{TT}\right ]\ .
\end{equation}

\noindent
These 4 responses can be separated by varying the (transverse)
polarization $\epsilon$ of the virtual photon as well as the
polarizations of the electron $(P_e)$ and proton ($P_z$ parallel and
$P_x$ orthogonal to the virtual photon, in the scattering plane). The
relations between the virtual photon cross sections and the quark spin
structure functions $g_1$ and $g_2$ can be read off the following
equations, which define a possible generalization of the GDH integral
and the BC sum rule,

\begin{eqnarray}
\label{DDeq5.10}
I_1(Q^2) & = & \frac{2m^2}{Q^2}\int\,g_1\ dx  
      =   \frac{m^2}{2\pi e^2}\int \frac{(1-x)}{\nu}\left [\sigma_{1/2}-
          \sigma_{3/2}+2\frac{Q}{\nu}\sigma'_{LT}\right ] d\nu\ , \\ 
I_2(Q^2) & = & \frac{2m^2}{Q^2}\int\,g_2\ dx\ 
      =   \frac{m^2}{2\pi e^2}\int \frac{(1-x)}{\nu}\left [-\sigma_{1/2}+
          \sigma_{3/2}+2\frac{\nu}{Q}\sigma'_{LT}\right ] d\nu\ .\nonumber 
\end{eqnarray}

\noindent
The integrals run over the Bjorken variable $x = Q^2/2m \nu$ from $0$
to inelastic threshold $x_{thr} < 1$, and over the lab energy
$\nu$ from inelastic threshold $\nu_{thr}$ to infinity.  In the real
photon limit the transverse-longitudinal function $\sigma'_{LT}$ does
not contribute to $I_1$, which is then given by the GDH sum rule, $I_1
(0) = - \kappa^2/4$. In the limit of large $Q^2$, the quark structure
functions are supposed to scale, $g_1 (x, Q^2) \rightarrow g_1 (x)$,
and $I_1 \rightarrow 2m \Gamma_1/Q^2$ with $\Gamma_1$=const.  In the
case of the proton, DIS scattering experiments have established that
$\Gamma_1^p > 0$. Therefore, the integral $I_1$ starts at the (large)
negative GDH value for $Q^2 = 0$ and approaches (small) positive
values for large $Q^2$.

The second integral of Eq.~(\ref{DDeq5.10}) can be expressed by the
Sachs form factors,

\beqn
\label{DDeq5.12}
I_2 (Q^2) = \frac{G_M (Q^2) (G_M (Q^2) - G_E (Q^2))}
                 {4 (1 + \frac{Q^2}{4 m^2})},
\eeqn

\noindent
which is the less familiar BC sum rule~\cite{DDBur70}.  The integral
vanishes as $Q^{-10}$ for large momentum transfer and approaches $I_2
(0) = \frac{1}{4} \mu \kappa$ for real photons.

There exists an extensive literature on model calculations of the GDH
sum rule and its generalizations. For reasons of brevity only a few
can be mentioned at this point. The evolution of the generalized GDH
integral was studied first by Anselmino in the framework of vector
meson dominance~\cite{DDAns89}. Burkert and Ioffe~\cite{DDBur92}
combined this model with the available information on
electroproduction cross sections in the resonance region, and Soffer
and Teryaev~\cite{DDSof93} pointed out the importance of the structure
functions $\sigma_{LT}'$ and $g_2$ at intermediate values of $Q^2$.
The integrals of Eq.~(\ref{DDeq5.10}) were recently calculated in the
framework of a gauge invariant and unitarized resonance model that
describes the existing electroproduction data for $\nu \le 1.1$ GeV
and $Q^2 < 3$GeV$^2$ quite well~\cite{DDDre98c}. In particular, the
model has the appropriate threshold behaviour as required by the LET,
agrees with the data for the $\Delta$ multipole $M_{1+} (Q^2)$ and
shows the right helicity structure for the second and third resonance
regions, the rapid change of the asymmetry from $\sigma_{3/2}$ to
$\sigma_{1/2}$ dominance with increasing $Q^2$. The calculations show
that even the small $\Delta$ multipoles $E_{1^+}$ and $S_{1^+}$, the
"bag deformation" in a simple quark model, could affect the integrals
of Eq.~(\ref{DDeq5.10}) at intermediate values of $Q^2$. These
multipoles are now being studied at Jefferson Lab, ELSA, MIT/Bates and
MAMI~\cite{DDSto98}.  In particular, the experiments at Jefferson Lab
indicate that $E_{1^+}/M_{1^+}$ remains of the order of $-5\%$
even at the highest values of $Q^2$, while perturbative QCD predicts
$100\%$ at sufficiently large $Q^2$.  The $Q^2$ evolution of the
N$\Delta$ transition form factors will also be addressed in an
effective chiral lagrangian framework by G.~C.~Gellas et
al.~\cite{DDGel98}.

Returning to the evolution of the generalized GDH integral $I_1(Q^2)$,
we find the zero-crossing of the integral at $Q^2_0 \approx 0.8$
GeV$^2$ if we include the one-pion contribution only. This value is
lowered to about 0.5 GeV$^2$ if we also add the $\eta$ production and,
in a crude model, the more-pion channels~\cite{DDDre98c}. In this
intermediate region the recent SLAC experiment~\cite{DDAbe97} yields
$I_1$ (0.5 GeV$^2) = 0.10 \pm 0.06$, while our result is only slightly
positive. There are at least two reasons for this (minor) deviation.
First, the strong dependence of the zero-crossing on the higher
production channels gives rise to uncertainties in our model. Second,
the error of the experiment is only the statistical one, and
systematical errors are estimated to be of equal size. In particular
the lack of data points near the $\Delta$ resonance could easily lead
to sizeable systematical errors.

The evolution of the generalized GDH integral is of considerable
current interest. Its dependence on $Q^2$ describes the transition
between the resonance dominated coherent process for small virtuality
and the incoherent process of DIS off the constituents at large $Q^2$.
Several experiments are planned and partially scheduled at Jefferson
Lab to explore the spin structure of the nucleon in the transition
region~\cite{DDBur91}. Their outcome will be quite invaluable for our
understanding of the nonperturbative phase of QCD.



\begin{thebibliography}{99}
\bibitem{DDArn81} R. G. Arnold et al., \Journal{\PRC}{23}{363}{1981}.
\bibitem{DDSch98} See contributions of H. Schmieden and P. Grabmayr
                  in these proceedings.
\bibitem{DDBal98} See contribution of R. Baldini in these proceedings.
\bibitem{DDDub98} See contributions of S. Dubnicka, 
                  A.-Z.~Dubnickov\'{a} and P. Weisenpacher
                  in these proceedings.
\bibitem{DDBij98} See contributions of R. Bijker and A. Leviatan, and
                  of S.~Boffi, P.~Demetriou, M.~Radici,
                  and R.~F.~Wagenbrunn in these proceedings.
\bibitem{DDWag98} See contributions of G. Wagner and A.~Buchmann
                  in these proceedings.
\bibitem{DDHab98} See contribution of H.~Haberzettl in these proceedings.
\bibitem{DDEsc98} See contribution of I.~Eschrich in these proceedings.
\bibitem{DDMue97} B. Mueller et al. (SAMPLE Collaboration), \Journal{\PRL}
                  {78}{3824}{1997}.
\bibitem{DDPit98} See contributions of M.~Pitt and R.~Michaels
                  in these proceedings.
\bibitem{DDHem98} See contribution of T.~Hemmert in these proceedings.
\bibitem{DDDre98a}D. Drechsel, G. Krein, and O.~Hanstein, \Journal{\PLB}
                  {420}{248}{1998}.
\bibitem{DDTon98} J. Tonnison et al., \Journal{\PRL}{80}{4382}{1998}.
\bibitem{DDLvo98} See contribution of A. I. L'vov in these proceedings;
                  A.~I.~L'vov and A.~M.~Nathan, submitted to {\it Phys.
                  Rev. Lett.} B, [nucl-th/980732].
\bibitem{DDPas98} B. Pasquini, in proceedings of the Joint
                  ECT$^{\ast}$/JLab Workshop on ``N$^{\ast}$~physics and
                  non-perturbative QCD'', Trento (1998)
\bibitem{DDGui95} P. A. M. Guichon, G. Q. Liu, and A.~W.~Thomas, \Journal
                  {\NPA}{591}{606}{1995}.
\bibitem{DDDre97} D. Drechsel et al., \Journal{\PRC}{57}{941}{1998}
                  and references given therein.
\bibitem{DDVan97} M. Vanderhaeghen, \Journal{\PLB}{402}{243}{1997}.
\bibitem{DDdHo98} See contributions of N.~d'Hose and J.~Friedrich
                  in these proceedings.
\bibitem{DDFil98} See contribution of L.~V.~Fil'kov and V.~L.~Kashevarov
                  in these proceedings.
\bibitem{DDMoi98} See contribution of M. A. Moinester in these proceedings.
\bibitem{DDBal60} A. M. Baldin, {\it Nucl. Phys.} {\bf 18}, 310 (1960).
\bibitem{DDGer65} S. B. Gerasimov, {\it Yad. Fiz.} {\bf 2}, 598 (1965),
                  [{\it Sov. J. Nucl. Phys.} {\bf 2}, 430 (1966)]; S.~D.~
                  Drell and A.~C.~Hearn, \Journal{\PRL}{16}{908}{1966}.
\bibitem{DDGel54} M. Gell-Mann, M. Goldberger, and W.~Thirring, \Journal
                  {\PREV}{95}{1612}{1954}.
\bibitem{DDKar73} I. Karliner, \Journal{\PRD}{7}{2717}{1973}.
\bibitem{DDSan94} A. M. Sandorfi, C. S. Whisnant, and M.~Khandaker,
                  \Journal{\PRD}{50}{R6681}{1994}.
\bibitem{DDDre98b}D. Drechsel and G. Krein, {\it Phys. Rev.} D (1998),
                  [hep-ph/9808230].
\bibitem{DDAre98} See contribution of J. Arends in these proceedings.
\bibitem{DDBur70} H. Burkhardt and W.~N.~Cottingham, {\it Ann. Phys. (N.Y.)}
                  {\bf 56}, 453 (1970).
\bibitem{DDAns89} M. Anselmino, B. L. Ioffe, and E.~Leader, {\it Yad. Phys.}
                  {\bf 49}, 136 (1989), [{\it Sov. J. Nucl. Phys.} {\bf 49},
                  136 (1989)].
\bibitem{DDBur92} V. D. Burkert and B. L. Ioffe, \Journal{\PLB}{296}{223}
                  {1992} and {\it JETP} {\bf 105}, 619 (1994).
\bibitem{DDSof93} J. Soffer and O. Teryaev, \Journal{\PRL}{70}{3373}{1993}
                  and \Journal{\PRD}{51}{25}{1995}.
\bibitem{DDDre98c}D. Drechsel, O. Hanstein, S.~S.~Kamalov, and L.~Tiator,
                  {\it Nucl. Phys.} A (1998) in print, [nucl-th/9807001];
                  D.~Drechsel, S.~S.~Kamalov, G.~Krein, and L.~Tiator,
                  to be published, [hep-ph/9810480].
\bibitem{DDSto98} See contributions of P. Stoler, R.~Gothe, M.~Distler,
                  and P.~Bartsch in these proceedings.
\bibitem{DDGel98} See contribution of G. C. Gellas, C.~N.~Ktorides,
                  G.~I.~Poulis, and T.~R.~Hemmert in these proceedings.
\bibitem{DDAbe97} K. Abe et al. (E143 Collaboration), \Journal{\PRL}{78}
                  {815}{1997} and {\it Phys. Rev.} D (1998), [hep-ph/9802357].
\bibitem{DDBur91} V. D. Burkert et al., CEBAF PR-91-23 (1991); S.~Kuhn et al.,
                  CEBAF~PR-93-09 (1993); Z.~E.~Meziani et al., CEBAF
                  PR-94-10 (1994); J.~P.~Chen et al., TJNAF PR-97-110 (1997).
\end{thebibliography}
\end{document}